\documentclass[aps,prx,twocolumn,longbibliography]{revtex4-2}

\usepackage{amsmath}
\usepackage{amssymb}
\usepackage{graphicx}
\usepackage{hyperref}
\usepackage{xcolor}
\usepackage{makecell}
\usepackage{booktabs}
\usepackage[section]{placeins}
\usepackage{amsthm}
\usepackage{multirow}

\makeatletter
\AtBeginDocument{%
  \expandafter\renewcommand\expandafter\subsection\expandafter{%
    \expandafter\@fb@secFB\subsection
  }%
}
\makeatother

\hypersetup{
    colorlinks=true,
    linkcolor=blue,
    citecolor=blue,
    urlcolor=blue
}

\begin{document}

\author{Jahan Claes}
\email{jahan@logiqal.com}
\affiliation{Logiqal Inc., Monmouth Junction, NJ 08852, USA}

\title{Cultivating $|T\rangle$ states on the surface code with only two-qubit gates}

\begin{abstract}
    High-fidelity $|T\rangle$ magic states are a key requirement for fault-tolerant quantum computing in 2D. It has generally been assumed that preparing high-fidelity $|T\rangle$ states requires noisy injection of $|T\rangle$ states followed by lengthy distillation routines. This assumption has been recently challenged by the introduction of \emph{cultivation}, in which careful state injection and postselection alone are used to prepare $|T\rangle$ states close to the fidelity required for quantum algorithms. Cultivation was originally proposed for the color code~\cite{gidneyMagicStateCultivation2024}, but can also be done on the $\mathbb{RP}^2$ code~\cite{chenEfficientMagicState2025}, a code similar to the surface code.

    In this work, we demonstrate how to cultivate $|T\rangle$ states directly on the surface code. Besides its simplicity compared to color or $\mathbb{RP}^2$ cultivation, surface code cultivation offers a number of advantages, including: (1) It is more directly compatible with neutral atom architectures than $\mathbb{RP}^2$ cultivation (2) Cultivated surface code states can be used in transversal CNOT gates with other surface codes, unlike color code states (3) Surface code cultivation can be done at any distance, unlike color and $\mathbb{RP}^2$ cultivation which requires odd distances. Under a standard depolarizing error model, our $d=3\ (d=4)\ (d=5)$ cultivation reaches an error rate of $1\cdot 10^{-6}\ (1\cdot 10^{-8})\ (2\cdot 10^{-9})$ and an acceptance rate of $34\%\ (6\%)\ (1\%)$, meeting or exceeding the fidelity of color and $\mathbb{RP}^2$ cultivation with comparable acceptance rates.
\end{abstract}

\maketitle

\section{Introduction}
Fault-tolerant quantum computation in 2D requires preparing high-fidelity \emph{magic states}, non-stabilizer states that can be used to teleport non-Clifford gates~\cite{gottesmanQuantumTeleportationUniversal1999,bravyiUniversalQuantumComputation2005,bravyiClassificationTopologicallyProtected2013,pastawskiFaulttolerantLogicalGates2015}. Historically, it was believed that preparing magic states required a two-step process: first, noisy magic states would be injected into the error-correcting code~\cite{liMagicStatesFidelity2015}; second, many noisy magic states would be distilled into a single high-fidelity magic state using protected Clifford operations~\cite{bravyiUniversalQuantumComputation2005,bravyiMagicStateDistillation2012,litinskiMagicStateDistillation2019}. However, it has recently been proposed that a more involved injection procedure using repeated error checks and postselection can generate much higher-fidelity $|T\rangle:=|0\rangle+e^{i\pi/4}|1\rangle$ magic states than previously thought possible, reducing or potentially eliminating the need for distillation. This postselection procedure is known as \textit{cultivation}~\cite{gidneyMagicStateCultivation2024}. 

In the original cultivation proposal~\cite{gidneyMagicStateCultivation2024}, as well as in earlier precursors to cultivation~\cite{itogawaEvenMoreEfficient2024,chamberlandVeryLowOverhead2020}, the authors injected $|T\rangle$ states into the color code~\cite{bombinTopologicalQuantumDistillation2006}. The color code posseses a transversal $H_{XY}:=(X+Y)/\sqrt{2}$ gate, allowing for double-checking the $|T\rangle$ state by fault-tolerantly measuring the logical $H_{XY}$ operator ($|T\rangle$ is the $+1$ eigenstate of $H_{XY}$, so measuring logical $H_{XY}$ checks that the logical state is correct). However, cultivating in the color code is not ideal, as many proposals for fault-tolerant quantum computing are based on the surface code~\cite{kitaevQuantumErrorCorrection1997,bravyiQuantumCodesLattice1998,litinskiGameSurfaceCodes2019,fowlerLowOverheadQuantum2019,fowlerSurfaceCodesPractical2012,gidneyHowFactor20482021,dennisTopologicalQuantumMemory2002}, which is easier to implement and decode. Ref~\cite{gidneyMagicStateCultivation2024} address this by grafting their cultivated color code into a larger matchable code with two surface-code-like boundaries capable of performing lattice surgery with other surface code qubits. However, the grafted code is somewhat unwieldy. It is not directly usable in architectures that use transversal CNOTs rather than lattice surgery, such as neutral atoms. It was also noted in Ref.~\cite[Fig. 16]{gidneyMagicStateCultivation2024} that the grafted code has a higher error rate than the surface code. It would thus be desirable to cultivate directly on the surface code.

Two recent proposals have made partial progress in surface code cultivation. Both of these proposals are tailored to neutral atoms, requiring the flexible connectivity enabled by reconfigurable qubits. First, Ref.~\cite{vakninMagicStateCultivation2025} proposed a method for cultivating two-qubit $|CX\rangle$ magic states on two patches of surface code. However, this method has three drawbacks: (1) it uses physical three-qubit CCX gates, which are more difficult to implement than two-qubit gates; (2) $|CX\rangle$ states cannot be used directly to teleport gates, but must be probabilistically converted into Toffoli states~\cite{dennis2001toward,gupta2024encoding}; (3) it produces $|CX\rangle$ states that are roughly $10^3\times$ noisier than the $|T\rangle$ states produced by color code cultivation. Second, Ref.~\cite{chenEfficientMagicState2025} proposed a method for cultivating $|T\rangle$ states on a variant of the surface code using only two-qubit gates; this method achieves similar fidelities to color-code cultivation and had an overall lower spacetime volume per $|T\rangle$ state, demonstrating the promise of surface code cultivation over color code cultivation. However, this method performs cultivation on the surface code defined on a non-orientable surface (the real projective plane, $\mathbb{RP}^2$, see also~\cite{kobayashi2024cross}) before escaping to the usual planar surface code.  One significant drawback in this approach is that the non-orientable boundary conditions of $\mathbb{RP}^2$ mean the cultivation circuit requires many long-distance connections that are difficult to implement, even in neutral atoms. In addition, low-depth unitary injection and expansion circuits are not known for the $\mathbb{RP}^2$ surface code, so that state injection and intermediate code growth must be done with measurement-based circuits which require mid-circuit measurements and feedback, significantly slowing the procedure in neutral atom platforms with slow measurements.

In this work, we combine ideas from $\mathbb{RP}^2$ cultivation with the mid-cycle trick~\cite{mcewenRelaxingHardwareRequirements2023,chenTransversalLogicalClifford2024} to directly cultivate $|T\rangle$ states on the planar surface code. Our overall approach is motivated by the constraints of neutral atom systems: We use only two-qubit gates, we restrict our qubit movements to rigid translations of a small number of fixed arrays of qubits, and we avoid mid-circuit measurement. Sticking to these restrictions means we do not necessarily use the most effective possible circuits in terms of depths, gate counts, or error rates, but nonetheless achieve fidelities competitive with color and $\mathbb{RP}^2$ cultivation protocols. 

We note that cultivation can be done at different distances, where increasing the distance increases the fidelity but decreases the success rate. Previous cultivation approaches were restricted to occur at odd $d$, since the color code cannot have even distance and the $\mathbb{RP}^2$ code is not fold-dual at even distance. Surface code cultivation, by contrast, can be done at any distance, allowing finer-grained trade-offs between fidelity and success rate.

\section{Results}

We begin with an overview of our approach to cultivation and a presentation of numerical results; we explain each stage of cultivation in more detail in the next section. 

\begin{figure*}[htb]
\includegraphics[width=\textwidth]{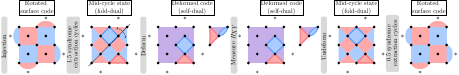}
\includegraphics[width=\textwidth]{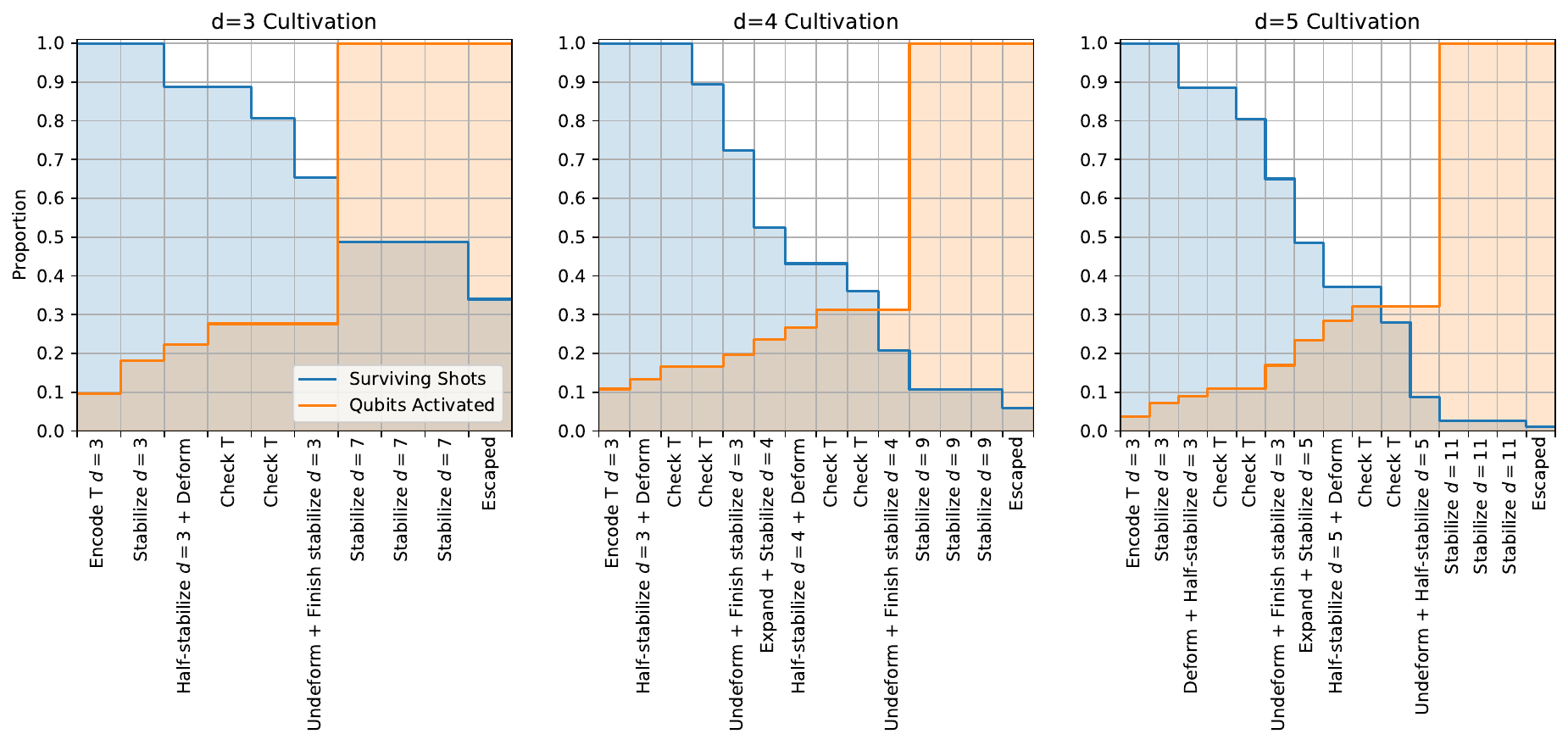}
\caption{Top: A sketch of distance-3 cultivation. We begin with a magic state injected into the rotated surface code. We then perform a full round of syndrome extraction, followed by another half-round of syndrome extraction to get the code into the mid-cycle state. The mid-cycle state is fold-dual across the diagonal line shown. We then use ancilla qubits and a short quantum circuit to deform the mid-cycle state to a self-dual code. Note that the the deformation maps the highlighted $Z$ ($X$) stabilizers of the mid-cycle state above (below) the fold to stabilizers below the fold; similarly, it maps $Z$ ($X$) stabilizers below (above) the fold to stabilizers above the fold, making the code self-dual. We measure $H_{XY}$ using an ancilla cat state (not shown), and reverse the deformation to return us to the mid-cycle state. We then complete the second half of the syndrome extraction cycle. From here, we can either grow the code to a higher distance and repeat the cultivation, or escape, depending on the desired fidelity of the final magic state.
\protect\\\hspace*{2em} Bottom: Timelines demonstrating each stage of $d=3,4,5$ cultivation. We grow the code (orange) while postselecting (blue). The expected spacetime volume is given by the integral of the product of these two curves, divided by the overall success rate. Note that this estimate of the spacetime volume is somewhat arbitrary, and was chosen only to enable comparisons with color code cultivation.}
\label{fig:CultivationSummary}
\end{figure*}

Cultivating $|T\rangle$ states on the planar surface code, just as in cultivation on the color code or $\mathbb{RP}^2$ surface code~\cite{gidneyMagicStateCultivation2024,chenEfficientMagicState2025}, proceeds in three stages:
\begin{enumerate}
    \item Injection: We prepare a magic state in a distance-3 surface code.
    \item Cultivation: We repeatedly measure the stabilizers and the logical $H_{XY}$ operator of the code, possibly growing the code to a higher-distance during this process. Measuring $H_{XY}$ verifies that we are in the logical $|T\rangle$ state, as this state is the $+1$ eigenstate of $H_{XY}$. In this stage, we postselect on any incorrect measurement outcomes, discarding the attempt.
    \item Escape: We grow the surface code to a larger distance surface code that is capable of protecting the logical information to the desired fidelity. In this stage, we postselect on the complementary gap~\cite{gidneyYokedSurfaceCodes2023}.
\end{enumerate}
By repeatedly measuring and postselecting at increasing distances in step 2, we generate injection circuits where all errors of weight up to $(d-1)$ are flagged and can be postselected out.

We illustrate the injection/cultivation stages of our distance-3 cultivation procedure in Fig.~\ref{fig:CultivationSummary}. For distance-5 cultivation, we run the distance-3 cultivation, then grow the surface code from $d=3$ to $d=5$ and repeat the cultivation procedure at $d=5$. For distance-4 cultivation, we run a modified distance-3 cultivation in which we only measure the stabilizers once, then grow the surface code from $d=3$ to $d=4$ and repeat the cultivation procedure at $d=4$ (now measuring the $d=4$ stabilizers twice). We note that, just in the case of color code cultivation, in order to actually flag all errors of weight less than $d=4$ or $d=5$, we actually need to measure the stabilizers at $d=4,5$ more than twice; however, just as in color code cultivation, we find that only measuring the stabilizers twice results in an improved acceptance rate with negligible impact on the fidelity.

For injection, we use a unitary circuit to prepare the distance-3 magic state. Similar to Ref.~\cite{gidneyMagicStateCultivation2024} we design an injection circuit where the $T$ rotation occurs midway through. This construction allows the circuit to detect single-qubit $X$ errors and ensures the only distance-1 errors in the injection circuit are $Z$ errors at the spacetime location of the $T$ gate. The unitary encoding circuit allows us to avoid the mid-circuit measurement and feedback required in Ref.~\cite{chenEfficientMagicState2025}.

In the cultivation stage, we need to both measure the stabilizers and measure the logical $H_{XY}$ operator. To measure the logical $H_{XY}$ operator, we want to deform the surface code into a \emph{self-dual} code where the $X$ and $Z$ stabilizers and logical operators exactly overlap and thus $H_{XY}$ is transversal. Ref.~\cite{chenEfficientMagicState2025} demonstrated that is it straightforward to deform \emph{fold-dual} code, where the $X$ and $Z$ stabilizers and logical operators are related by reflection across a fold line, into a self-dual code using ancilla qubits. However, the rotated surface code is not fold-dual, so we cannot directly use this method. Fortunately, midway through the syndrome extraction circuit, the mid-cycle state of the rotated surface code is the \emph{unrotated} surface code, which \emph{is} fold-dual across the diagonal. Therefore, we choose to measure $H_{XY}$ halfway through the syndrome extraction cycle, by deforming the mid-cycle state into a self-dual code. To deform the mid-cycle state into a self-dual code, we use a simpler method than Ref.~\cite{chenEfficientMagicState2025}, as it is easier to preserve the distance of the of the mid-cycle state than the $\mathbb{RP}^2$ code. We then optionally grow the code from $d=3$ to $d=4,5$ using another unitary circuit, again avoiding the mid-circuit measurement and feedback in Ref.~\cite{chenEfficientMagicState2025}.

Finally, our escape state is more straightforward than either of Refs.~\cite{gidneyMagicStateCultivation2024,chenEfficientMagicState2025}; we simply grow the surface code directly into a larger surface code by initializing new qubits in $|+\rangle$ and $|0\rangle$ and measuring stabilizers of the larger code, as in any other surface code injection protocol~\cite{liMagicStatesFidelity2015,litinskiGameSurfaceCodes2019,horsman2012surface,fowlerSurfaceCodesPractical2012}. After measuring the larger code stabilizers several times, we can compute the complimentary gap to estimate how confident the decoder is in our result~\cite{gidneyYokedSurfaceCodes2023,gidneyMagicStateCultivation2024}. By varying the cutoff for accepting the state, we trade off between the fidelity of the final magic state and the number of retries required to accept the state.

\begin{table*}[htb]
    \centering
    \begin{tabular}{c|c|c|c|c|c|c|c|c|}
    &\makecell{Cultivation \\protocol} & \makecell{Noise\\model} & \makecell{Error rate\\ (ungrown)} & \makecell{Discard rate \\(ungrown) } & \makecell{Error rate\\ (end-to-end)} & \makecell{Discard rate \\(end-to-end)}& \makecell{Final\\ distance} & \makecell{Spacetime volume\\(qubit cycles)}\\\Xhline{3\arrayrulewidth}
    \multirow{4}{*}{\rotatebox[origin=c]{90}{Prev work}}&$d=3$ color~\cite{gidneyMagicStateCultivation2024}& \multirow{4}{*}{\makecell{Uniform\\depolarizing}} & $6\cdot 10^{-7}$ & $35\%$ & $3\cdot 10^{-6}$ & $80\%$&$15$&$8100$\\
    &$d=5$ color~\cite{gidneyMagicStateCultivation2024}& &$6\cdot 10^{-10}$&$85\%$&$2\cdot 10^{-9} $&$99\%$&$15$&$57000$\\
    &$d=3$ $\mathbb{RP}^2$~\cite{chenEfficientMagicState2025}& & $2\cdot 10^{-6}$&$49\%$&$3\cdot 10^{-6}$&$58\%$&$7$&$900$\\
    &$d=5$ $\mathbb{RP}^2$~\cite{chenEfficientMagicState2025}& &$7\cdot 10^{-10}$&$91\%$& $2\cdot 10^{-9}$&$93\%$&11&$6200$\\\hline
    \multirow{6}{*}{\rotatebox[origin=c]{90}{This work}}&$d=3$ surface& \multirow{3}{*}{\makecell{Uniform \\depolarizing}}&$1\cdot 10^{-6}$&$54\%$&$1\cdot 10^{-6}$&$66\%$&$7$&$900$\\
    &$d=4$ surface& &$9\cdot 10^{-9}$&$90\%$&$1\cdot 10^{-8}$&$94\%$&$9$&$4400$\\
    &$d=5$ surface& &$1\cdot 10^{-9}$&$98\%$&$2\cdot 10^{-9}$&$99\%$&$11$&$24600$\\\cline{2-9}
    &$d=3$ surface& \multirow{3}{*}{\makecell{Uniform \\depolarizing\\without idle}}&$4\cdot 10^{-8}$&$17\%$& $9\cdot 10^{-8}$&$33\%$&7&$650$\\
    &$d=4$ surface& &$2\cdot 10^{-11}$&$39\%$&-&-&-&-\\
    &$d=5$ surface& &$5\cdot 10^{-12}$&$54\%$&-&-&-&-\\\hline
    \end{tabular}
    \caption{Comparison of error rates, discard rates, and spacetime volumes for previous cultivation protocols versus this work, for a physical depolarizing error rate $p=10^{-3}$. The ungrown error rates (without escape) were estimated by enumerating all logical errors up do to distance $5$. Note that~\cite{gidneyMagicStateCultivation2024} reported the error rate as double the simulated $S|+\rangle$ cultivation error rate, to account for the difference between cultivating $|T\rangle$ and $S|+\rangle$, while~\cite{chenEfficientMagicState2025} reported the error rate as the simulated $S|+\rangle$ error rate (so the reported error rates in~\cite{chenEfficientMagicState2025} are smaller than~\cite{gidneyMagicStateCultivation2024} by a factor of two). We have made all numbers in this chart consist with the doubling convention of~\cite{gidneyMagicStateCultivation2024}.} 
    \label{tab:CultivationNumericsComparison}
\end{table*}
\begin{figure*}[htb]
    \includegraphics[width=\textwidth]{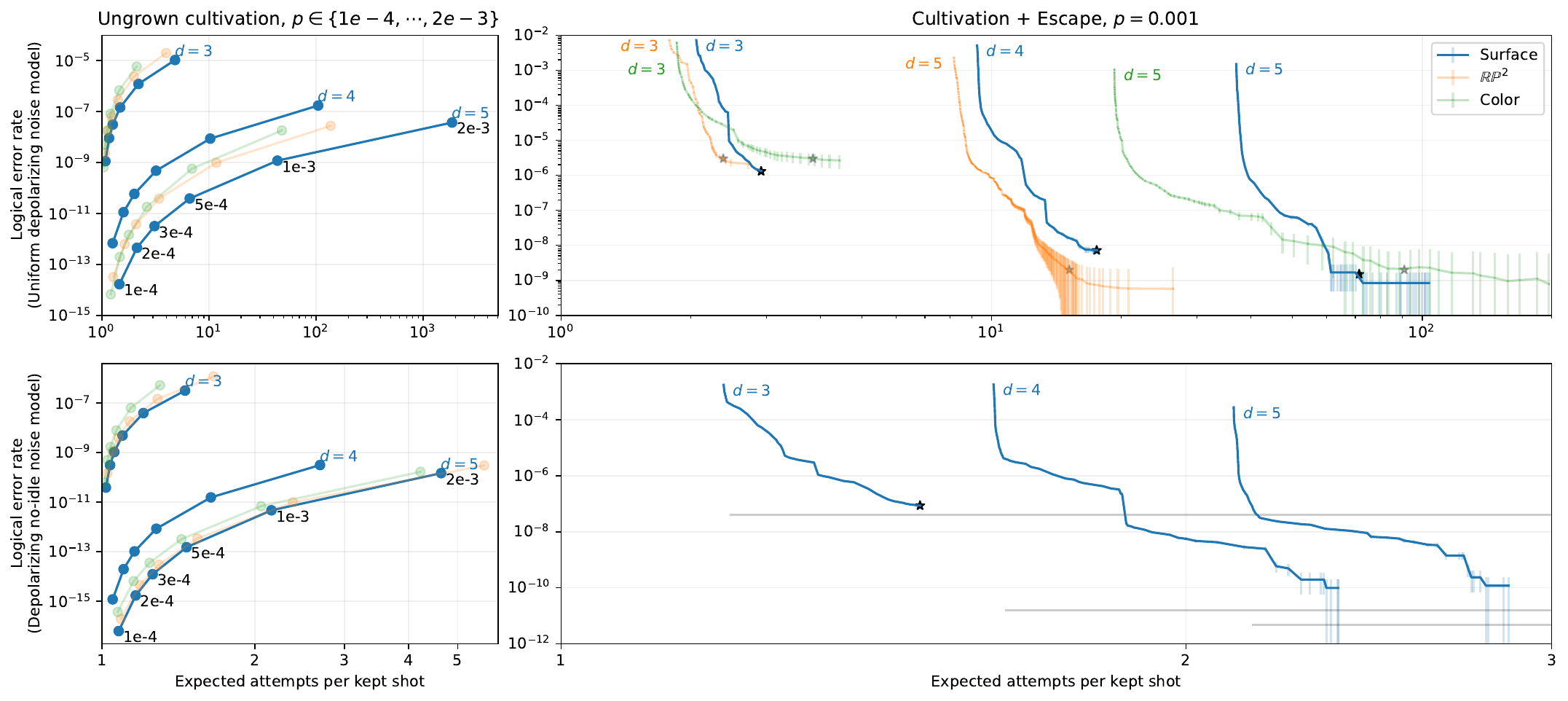}
    \caption{The error and discard rates of surface code cultivation, compared to previous work on color~\cite{gidneyMagicStateCultivation2024} and $\mathbb{RP}^2$~\cite{chenEfficientMagicState2025} cultivation. We plot two different error models, the uniform depolarizing model studied in~\cite{gidneyMagicStateCultivation2024,chenEfficientMagicState2025} (top row) and a depolarizing model with idle errors removed relevant to neutral atom qubits (bottom row). The left column shows the error/discard rates for cultivation without escape for various values of the physical error rate $p$, estimated by enumerating all errors of weight $\leq 5$; the right column shows the error/discard rates of cultivation plus escape for $p=0.001$, estimated using STIM~\cite{gidneyStimFastStabilizer2021} and PyMatching~\cite{higgottPyMatchingPythonPackage2021}. Stars in the right column denote the points used for Table~\ref{tab:CultivationNumericsComparison}.
    \protect\\\hspace*{2em}Note that the results in the bottom right are not directly comparable with previous work, as previous cultivation protocols did not provide simulation results for this error model. The $d=4,5$ simulations establish we can reach a $10^{-10}$ error rate at fairly low discard rate, but we do not have sufficient data to determine whether a higher gap cutoff could result in even lower error rates. The horizonal lines in the bottom right figure represent the ungrown logical error rates taken from the bottom left figure, which serves as a lower bound for the achievable logical error rate.
    \protect\\\hspace*{2em}For all panels, we double the simulated error rate as in~\cite{gidneyMagicStateCultivation2024}.}
    \label{fig:CultivationNumerics}
\end{figure*}

We report numerical simulations of our procedure compared to previous procedures in Fig.~\ref{fig:CultivationNumerics} and Table~\ref{tab:CultivationNumericsComparison}, using a standard depolarizing error model with $p=0.001$ (see Appendix for details). To enable efficient Clifford simulations, we follow Refs.~\cite{gidneyMagicStateCultivation2024,chenEfficientMagicState2025} and replace the $|T\rangle$ state with the Clifford state $S|+\rangle$ in our simulations. As in Ref.~\cite{gidneyMagicStateCultivation2024}, we report our error rate as double the error rate given by the $S|+\rangle$ simulation, to roughly account for an empirically observed difference between $S|+\rangle$ and $|T\rangle$ cultivation. Compared to previous cultivation methods, our cultivation achieves lower (for $d=3$) or similar (for $d=5$) logical error rates than $\mathbb{RP}^2$ and color code cultivation, while having a discard rate and spacetime volume between that of color and $\mathbb{RP}^2$ cultivation. In addition, our $d=4$ cultivation has a fidelity suprisingly close to $d=5$ cultivation at much lower discard rates. Due to limited computational resources, we do not simulate $d=6$ cultivation, but we expect $d=6$ surface code cultivation to be a much more feasible route to cultivate states with fidelity $<10^{-10}$ than $d=7$ color or $\mathbb{RP}^2$ cultivation. Our simulations are all performed with three rounds of error correction at the final distance before computing the complementary gap; we have tested that our results do not notably change when instead using five rounds of error correction, indicating both that we are not leaving potential gap information on the table by using too few error correction rounds at the final distance, and that our final distances are sufficiently large to protect our logical states to the desired fidelity with a negligible amount of postselection.

We also include numerical simulations of our cultivation circuits under a modified depolarizing error model without idle errors, which may be relevant for neutral atom qubits with extremely long idle times. In this error model, we can establish that the ungrown logical error rates (before the escape stage) are 2 to 3 orders of magnitude lower, and the ungrown discard rates are a factor of 2 to 3 lower. However, establishing the error and discard rates of cultivation plus escape is prohibitively expensive for $d=4,5$. We note that for previous simulations, the escape stage approximately doubles or triples the error rate, so we may conjecture that the $d=4,5$ surface codes still maintain errors $\sim 10^{-11}$ under this error model; however, our simulations can only establish error rates of $<10^{-10}$. When comparing to color or $\mathbb{RP}^2$ cultivation, we do not have simulations of cultivation plus escape, as previous work did not consider this error model. However, we note that unlike in the case of the uniform depolarizing error model, the performance of all three ungrown $d=5$ cultivation circuits are quite similar, suggesting that $d=5$ surface code cultivation may match $\mathbb{RP}^2$ cultivation under this error model. While we do not highlight these numbers in our table, as they are not relevant to the comparison with prior work, it suggests that the performance of cultivation on a given platform is highly sensitive to the details of the error model.

We end with two notes of caution on comparing the results of this paper to previous cultivation protocols. First, the spacetime volume figures in Fig.~\ref{fig:CultivationSummary} and Tab.~\ref{tab:CultivationNumericsComparison} should not be taken too seriously. Our division of our circuit into ``cycles'' is fairly arbitrary, and done only to allow for rough comparison to previous work. Notably, our circuit contains operations like code deformation that do not exist in the color code cultivation circuit. In addition, the number of active qubits (the orange curve in Fig.~\ref{fig:CultivationSummary}) depends on how we choose to rearrange and recycle qubits. Finally, we note that spacetime volume computed via the blue curves in~\ref{fig:CultivationSummary} assumes that we pause after every cycle to complete the measurement before deciding whether to continue the cultivation attempt; this was also assumed for the $\mathbb{RP}^2$ spacetime volume in Ref.~\cite{chenEfficientMagicState2025}. However, for neutral atoms qubits with slow measurements, it may be more efficient to batch measurements. Indeed, one virtue of our approach over Ref.~\cite{chenEfficientMagicState2025} is that no part of our circuit requires feed-foreward from a previous measurement outcome, meaning we do not need to be slowed down by slow measurement times. We thus caution that the spacetime volume is not necessarily a meaningful metric of comparison, and more hardware-specific measures will be needed in the future.

Second, we note that both the uniform depolarizing and the depolarizing no-idle error models likely do not capture the relevant features of neutral atom qubits. A more accurate error model would include noise induced by qubit movement and transfers between traps, in which the relative performance of the $\mathbb{RP}^2$ circuits would likely degrade but the performance of the color circuits may improve somewhat. These simulations also do not account for qubit leakage, which becomes increasingly damaging as the circuit depth increases. On the other hand, we also do not currently take advantage of the structure of noise available in certain in neutral atoms, such as erasure bias~\cite{wuErasureConversionFaulttolerant2022,sahayHighthresholdCodesNeutralatom2023,maHighfidelityGatesMidcircuit2023,senoo2025high,muniz2025high}, which may improve performance~\cite{zhangLeveragingErasureErrors2025,jacobyMagicStateInjection2025}. As a rough approximation, if a fraction $R_e$ of errors are erasures, the leading order error rate of the ungrown patch will go from $Cp^{d}$ to $C((1-R_e) p)^{d}$ for some constant $C$, meaning that a realistic $90\%$ erasure fraction~\cite{wuErasureConversionFaulttolerant2022} could improve the logical error rate of $d=5$ cultivation by a factor of $10^5$ (see also~\cite{jacobyMagicStateInjection2025}).

\section{Constructing the cultivation circuit}

We begin by noting that $|T\rangle$ is the unique $+1$ eigenstate of the operator $H_{XY}:=(X+Y)/\sqrt{2}$. Our goal is to inject $|T\rangle$ with as high a fidelity as possible, then alternate measuring the stabilizers with fault-tolerantly measuring the logical $H_{XY}$ operator to ensure we have successfully prepared the $|T\rangle$ state. We can do this measurement repeatedly at increasing code distances, if desired. Measuring $H_{XY}$ is the primary challenge, since the surface code does not posses a transversal $H_{XY}$ gate.

\subsection{Injection and code growth}

\begin{figure*}[htb]
    \includegraphics[width=\textwidth]{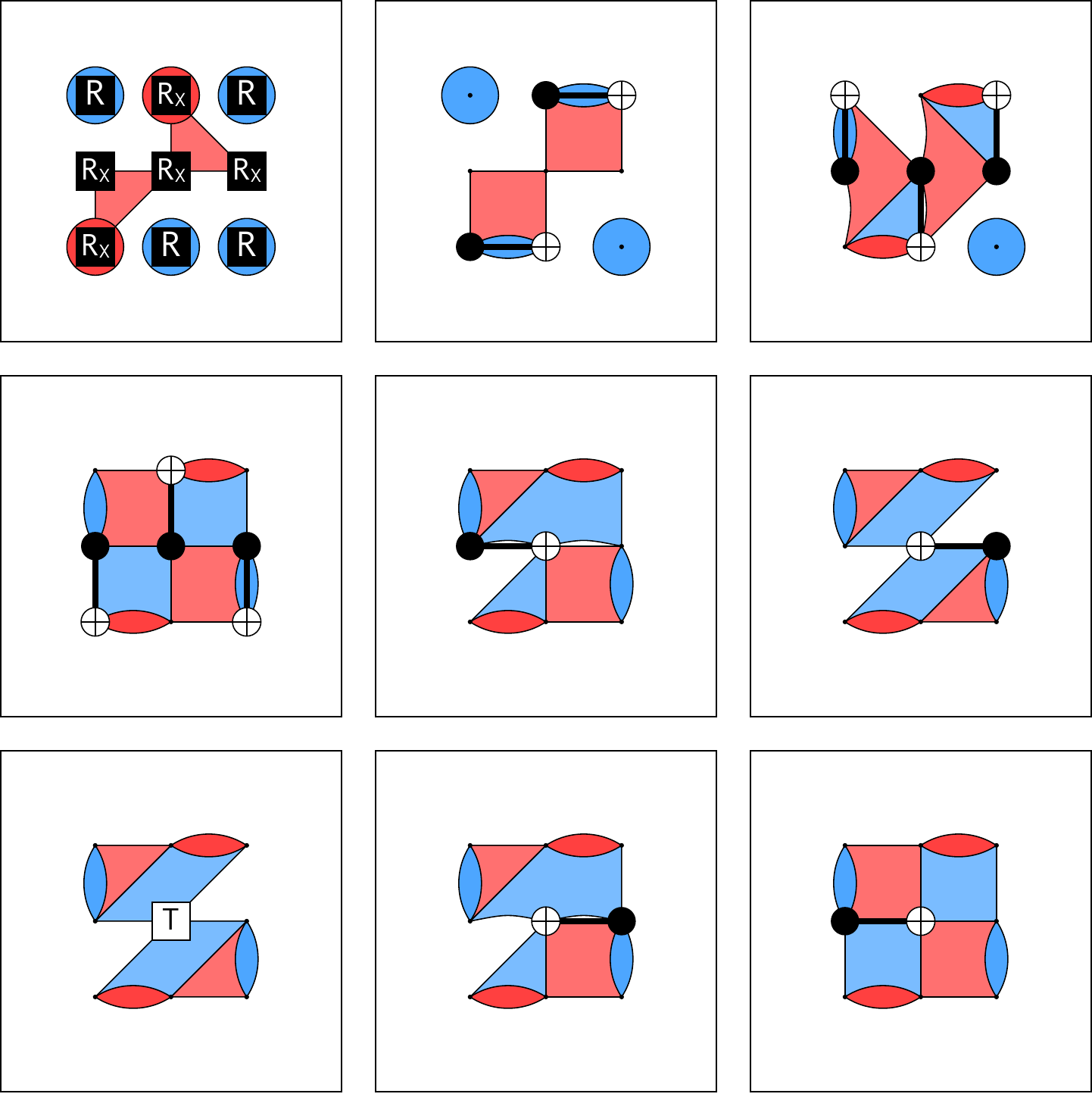}
    \caption{The unitary circuit we use to inject a magic state into the surface code, adapted from similar circuits in~\cite{satzingerRealizingTopologicallyOrdered2021}. We illustrate the eight initial stabilizers that evolve to be the stabilizers of the $d=3$ surface code. The logical $X$ operator begins as a single physical $X$ on the center qubit (not shown). This means that at initialization, a single-qubit error flipping the $X$ logical operator would also flip the two triangular $X$ stabilizers, making this error detectable. Furthermore, the $T$ rotation occurs with two $Z$ stabilizers touching it, meaning that any single-qubit $X$ or $Y$ error at this location is detectable. Overall, the only undetectable single-qubit errors in this circuit are $Z$ errors at the spacetime location of the $T$ gate.}
    \label{fig:InjectionCircuit}
\end{figure*}

Following \cite{gidneyMagicStateCultivation2024}, we use a unitary injection circuit to prepare the $|T\rangle$ state, adapted from the unitary injection circuit for $|0\rangle$ and $|+\rangle$ given in~\cite{satzingerRealizingTopologicallyOrdered2021}. Our circuit is illustrated in Fig.~\ref{fig:InjectionCircuit}. This circuit detects all distance-$1$ $X$ errors, and the only undetectable distance-$1$ $Z$ errors are those at the spacetime location of the $T$ gate. This is similar to the injection circuit in Ref.~\cite{gidneyMagicStateCultivation2024} which is similarly protected against distance-$1$ errors, and stands in contrast to the measurement-based $\mathbb{RP}^2$ injection circuit in Ref.~\cite{chenEfficientMagicState2025} which has a greater number of distance-$1$ errors (for example, the $\mathbb{RP}^2$ circuit is exposed to single-qubit errors at initialization). In addition, the measurement-based $\mathbb{RP}^2$ injection circuit requires mid-circuit measurement and feedback to project the code into the $+1$ eigenstate of all stabilizers, which we avoid with our unitary initialization.
\begin{figure*}[htb]
    \includegraphics[width=\textwidth]{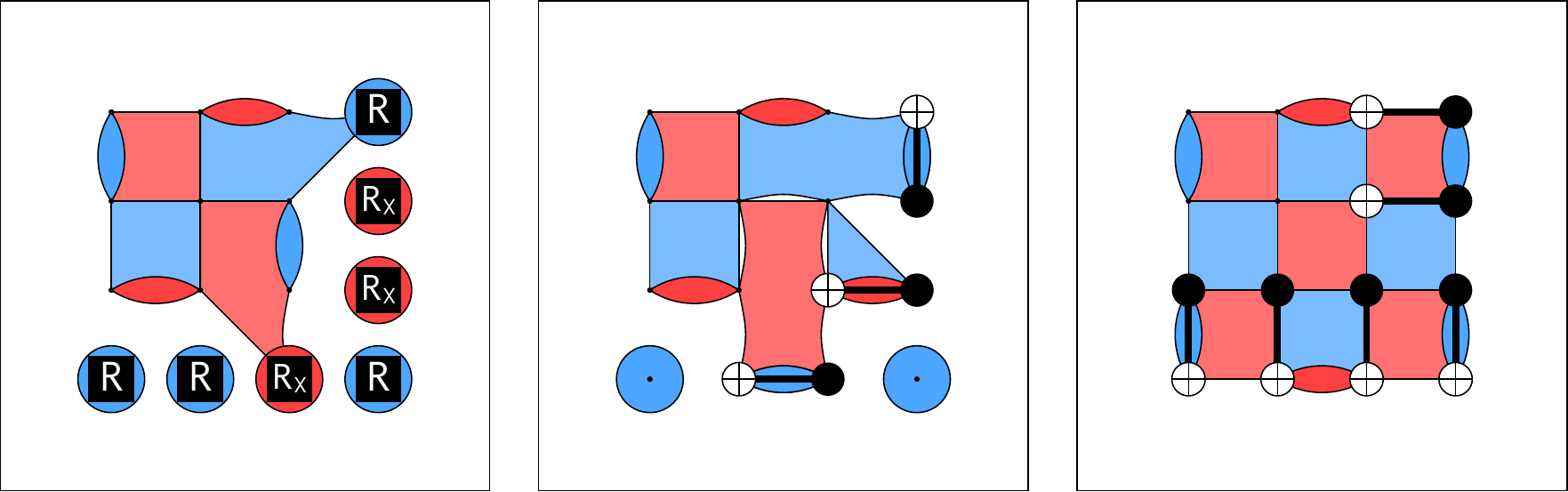}
    \
    
    \includegraphics[width=\textwidth]{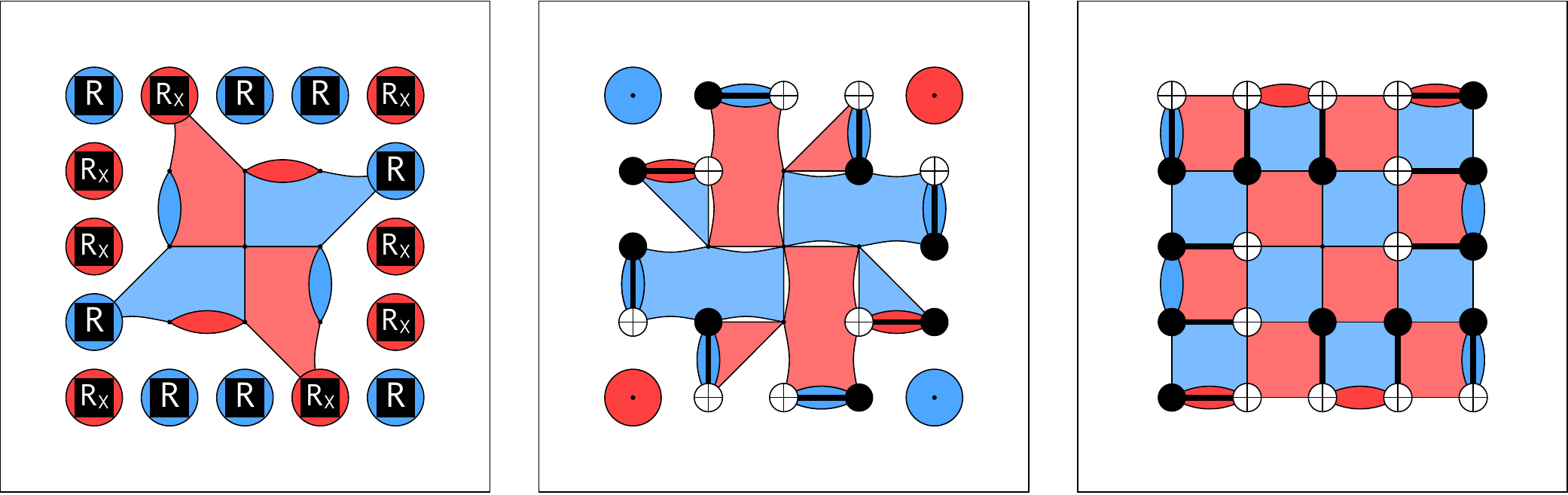}
    \caption{Unitary circuits to grow the surface code from $d=3$ to $d=4$ (top) or $d=5$ (bottom). Modified versions of the low-depth injection circuits introduced in~\cite{claes2025lower}.}
    \label{fig:ExtensionCircuit}
\end{figure*}

To grow a $d=3$ surface code to a $d=4$ or $d=5$ surface code, we use the unitary growth circuits illustrated in Fig.~\ref{fig:ExtensionCircuit} which are modified versions of the low-depth circuits introduced in~\cite{claes2025lower}. Unitary growth also avoids the midcircuit measurement and feedback used to grow the $d=3$ $\mathbb{RP}^2$ code to a $d=5$ $\mathbb{RP}^2$ code in Ref.~\cite{chenEfficientMagicState2025}. 

\subsection{Measuring logical $H_{XY}$}

We will see below that it is relatively straightforward to fault-tolerantly measure the logical $H_{XY}$ operator in codes that admit a transversal $H_{XY}$ gate~\cite{itogawaEvenMoreEfficient2024,chamberlandVeryLowOverhead2020,gidneyMagicStateCultivation2024}. The surface code does not admit a transversal $H_{XY}$ gates, but self-dual codes do. In the same spirit as Ref.~\cite{chenEfficientMagicState2025}, we will deform our surface code into a self-dual code in order to measure the logical $H_{XY}$ operator.

\begin{figure*}[htb]
    \includegraphics[width=\textwidth]{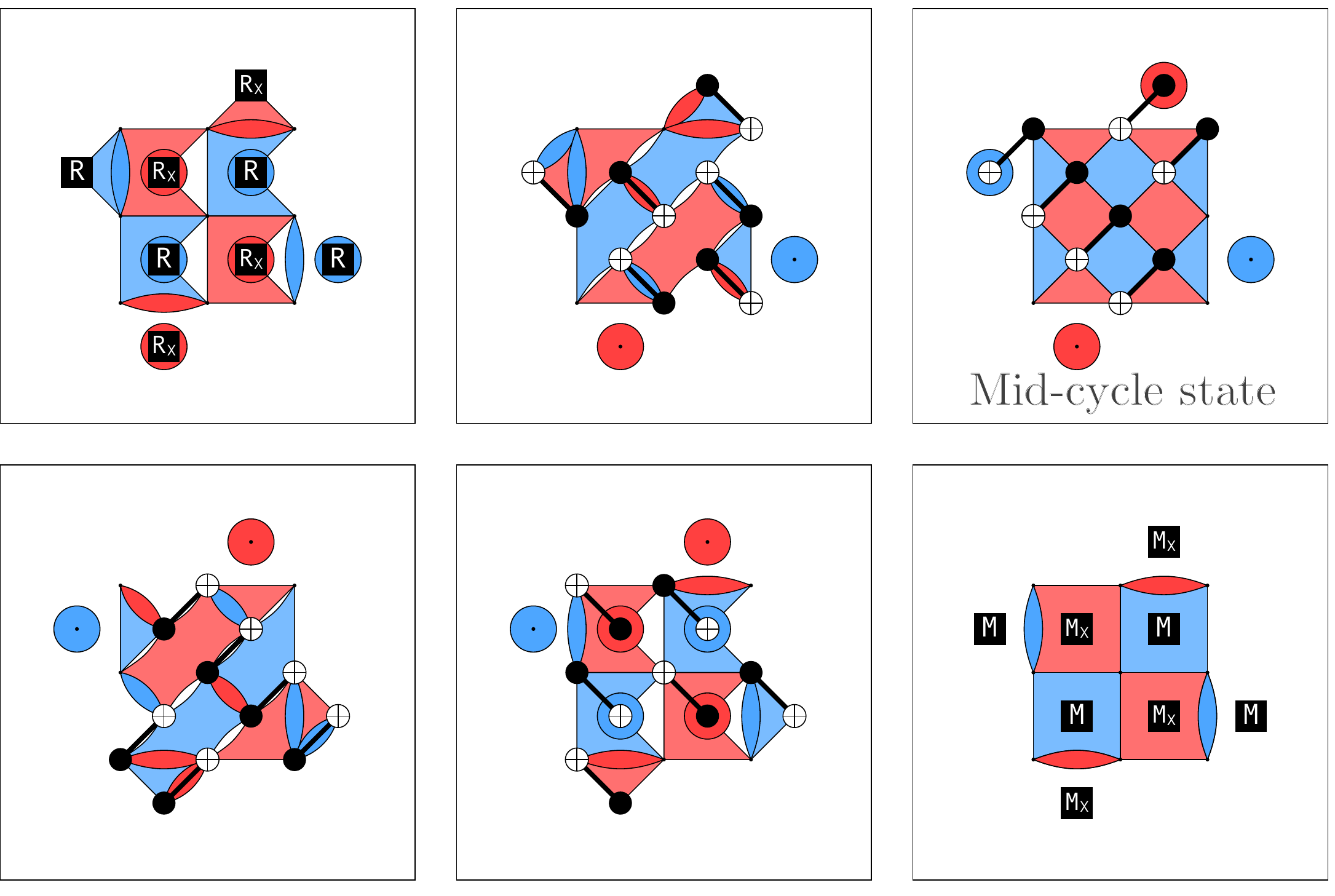}
    \caption{The evolution of stabilizers of a rotated surface code over a syndrome extraction cycle. We see that in the mid-cycle state, the stabilizers evolve to the stabilizers of the unrotated surface code, with the outer ancillas in product states~\cite{mcewenRelaxingHardwareRequirements2023,chenTransversalLogicalClifford2024}. Ignoring the trivial outer ancillas, the mid-cycle state is fold-dual across either diagonal line.}
    \label{fig:SyndromeCycle}
\end{figure*}

\subsubsection{Transversal $H_{XY}$ on self-dual codes}

It has long been known that the logical operators $H_{XY}$ and $iZH_{XY}=(X-Y)/\sqrt{2}$ can be implemented transversally in \emph{self-dual} CSS codes~\cite{bombinTopologicalQuantumDistillation2006,kubica2015universal}, where each $\otimes Z_i$ stabilizer is in one-to-one correspondence with a dual $\otimes X_i$ stabilizer supported on the same qubits (and the same for a basis of logical operators). Given a self-dual basis of logical operators, each operator is necessarily supported on an odd number of qubits $(2w+1)$, where in codes with multiple logical operators $w$ can depend on the particular logical operator. We claim that a transversal $H_{XY}$ gate applied to all the physical qubits applies a logical $H_{XY}$ to logical qubits with $w$ even or a logical $iZH_{XY}=(X-Y)/\sqrt{2}$ to logical qubits with $w$ odd. Conversely, a transversal $iZH_{XY}$ applies a logical $iZH_{XY}$ to logical qubits with $w$ even and a logical $H_{XY}$ to logical qubits with $w$ odd.

To demonstrate this claim, we first note that in the Heisenberg picture $H_{XY}$ sends $X\leftrightarrow Y$ and $Z\rightarrow-Z$. Similarly, $iZH_{XY}$ sends $X\leftrightarrow -Y$ and $Z\rightarrow -Z$. Thus, we need to show that the transversal $H_{XY}$ maps each set of self-dual logical Pauli operators as $X_L\leftrightarrow (-1)^w Y_L$ and $Z_L\rightarrow -Z_L$, while preserving each stabilizer.

We begin with the logical operators. $Z_L$ is given by a product of $(2w+1)$ physical $Z$ operators, $Z_L=\otimes^{2w+1}Z_i$. Since each physical $Z_i$ operator is sent to $-Z_i$ by transversal $H_{XY}$, $Z_L$ is sent to $(-1)^{2w+1}Z_L=-Z_L$. As for $X_L$ and $Y_L$, we note that $Y_L = i X_LZ_L=(-1)^w\otimes^{2w+1}Y_i$, so $H_{XY}$ sends $X_L\rightarrow \otimes^{2w+1}Y_i=(-1)^w Y_L$ and $Y_L\rightarrow (-1)^w\otimes^{2w+1}X_i=(-1)^w X_L$. Thus, for logicals with even $w$ (i.e., distance $5,9,13,...$) the transversal $H_{XY}$ acts as a logical $H_{XY}$, while for logicals with odd $w$ (i.e., distance $3,7,11,...$) it acts as a logical $iZ H_{XY}$.

We now turn to the stabilizers. In any self-dual code, every stabilizer has even weight. If we apply a transversal $H_{XY}$ operator to each qubit, it will send a $2w$-qubit stabilizer $\otimes^{2w} Z_i\rightarrow(-1)^{2w}\otimes^{2w} Z_i=\otimes^{2w} Z_i$. It will also send a $2w$-qubit stabilizer $\otimes^{2w} X_i$ to $(-1)^w \otimes^{2w} X_i \otimes^{2w} Z_i$, which is equal to $\otimes^{2w} X_i$ provided that we are in the $(-1)^w$ eigenstate of the $Z$ stabilizer. In other words, transversal $H_{XY}$ preserves the stabilizers, provided we are in a code state where each singly-even $Z$ stabilizer (weight $2$, $6$, $10$, etc) is $(-1)$ and each doubly-even stabilier (weight $4$, $8$, $12$, etc) is $(+1)$. As long as our stabilizers have been initialized in a known value, we can ensure the $Z$ stabilizers have the appropriate value by applying a layer of Pauli $X$ operators; we can also ensure that the stabilizers have the appropriate values by initializing some qubits in $|1\rangle$ rather than $|0\rangle$ in the injection and growth steps. Note that the fact that we need the $Z$ stabilizers to have certain values is why we prefer unitary initialization circuits, rather than measuremenent-based initialization circuits which project stabilizers into random values that are not known until after measurement.

We can similarly show that a transversal $iZH_{XY}$ operator acts as a logical $iZH_{XY}$ on logicals with $w$ even and a logical $H_{XY}$ on logicals with $w$ odd, with the same requirement on the value of the $Z$ stabilizers.

\subsubsection{Deforming to self-dual}

To measure the logical $H_{XY}$ operator, we want to deform the rotated surface code into a self-dual code with a short quantum circuit. To develop our deformation circuit, we use two facts. First, it was pointed out in Ref.~\cite{chenEfficientMagicState2025} that starting from a \emph{fold-dual} code, in which there exists some fold line such that each $\otimes X_i$ stabilizer is in one-to-one-correspondence with a $\otimes Z_i$ stabilizer across the fold (and the same for a basis of logical operators), it is relatively simple to deform into a self-dual code. Second, it is known that halfway through the syndrome extraction cycle of the rotated surface code, the mid-cycle state of the unrotated surface code plus its ancilla measurement qubits is the state of the unrotated surface code~\cite{mcewenRelaxingHardwareRequirements2023,chenTransversalLogicalClifford2024}, which is fold-dual across the diagonal line (see Fig.~\ref{fig:SyndromeCycle}). Overall, then, our strategy is to first transform the unrotated surface code into a fold-dual code at no cost (since we need to measure the stabilizers anyway), and then use a short quantum circuit to deform the fold-dual mid-cycle state into a self-dual code.

\begin{figure*}[htb]
    \includegraphics[width=\textwidth]{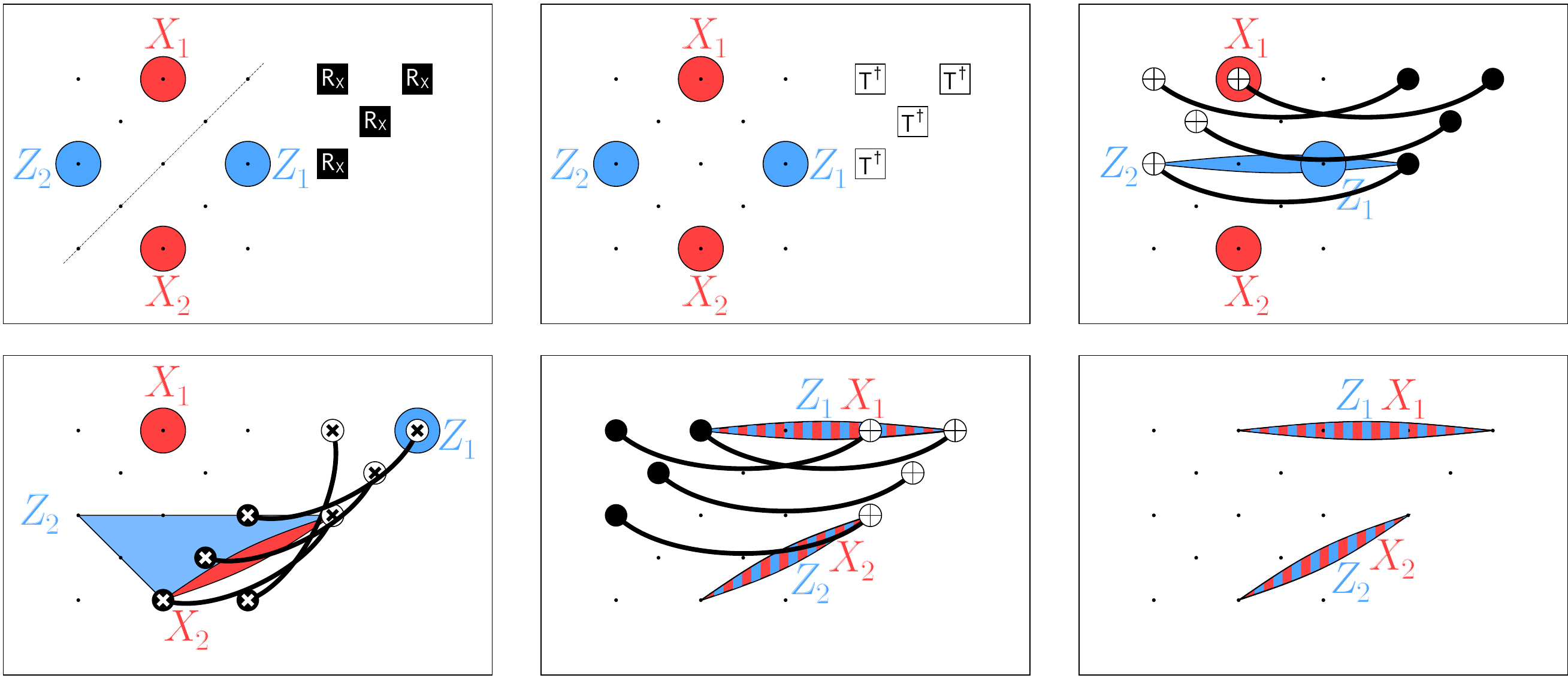}
    \caption{The deformation circuit takes us from the unrotated surface code to a self-dual code with a transversal $H_{XY}$. The gate denoted by black/white circles with $\times$s at the center is the CXSWAP gate which is equivalent to pair of CX gates, the first with the control on the white circle and the second with the control on the black circle. We illustrate two pairs of $X$ and $Z$ operators that are related by reflection across the fold. Our circuit takes $X$ operators above the fold and $Z$ operators below the fold to $XX$/$ZZ$ operators on the qubit above the fold and the matching qubit on the triangular ancilla patch. Similarly, it takes $Z$ operators above the fold and $X$ operators below the fold to $ZZ$/$XX$ operators on the qubit below the fold and the matching qubit on the triangular ancilla patch. The ancilla qubits are initialized in $T^\dagger|+\rangle$, making the logical degrees of freedom introduced by the ancilla qubits invariant under transversal $H_{XY}$. To undo the deformation, we run the circuit in reverse and measure the triangular ancilla patch in the $X$ basis.}
    \label{fig:DeformUndeform}
\end{figure*}

Our deformation circuit should take each single-qubit $X$ operator and its fold-dual $Z$ partner to overlapping $\otimes X$ and $\otimes Z$ operators. This is not possible do to without ancilla qubits; however, with a single ancilla for each pair of fold-dual qubits, we can arrange for the $X$ and $Z$ operators to overlap. We illustrate our circuit for $d=3$ in Fig.~\ref{fig:DeformUndeform}. Each pair of fold-dual qubits has a single ancilla qubit associated with it, and we have arranged the ancilla qubits into a triangular patch next to the surface code. Our circuit takes $X$ operators above the fold and $Z$ operators below the fold to $XX$/$ZZ$ operators on the qubit above the fold and the matching qubit on the triangular ancilla patch. Similarly, it takes $Z$ operators above the fold and $X$ operators below the fold to $ZZ$/$XX$ operators on the qubit below the fold and the matching qubit on the triangular ancilla patch. The result is that all $X$ and $Z$ stabilizers (and the $X$ and $Z$ logical operator of the surface code) are now overlapping and the code is thus self-dual. In terms of the distance $d$ of the surface code, the logical operators have weight $(2d-1)$, so that for odd $d$ the logical $H_{XY}$ is given by transversal $H_{XY}$, while for even $d$ it is given by transversal $iZH_{XY}$.

Note that the ancilla qubits introduce their own self-dual logical operators, since we have increased the number of qubits in the code but not increased the number of stabilizers. These additional logical operators must be initialized in a state that is invariant under the transversal $H_{XY}$/$iZH_{XY}$ operator, in order for this transversal operator to only act on the logical surface code degree of freedom. Each self-dual ancilla logical operator has weight three, so the transversal $H_{XY}$/$iZH_{XY}$ operator acts as a logical $iZH_{XY}$/$H_{XY}$ operator on these logical qubits. To ensure these degrees of freedom are invariant, we initialize the ancilla qubits in $|T\rangle$ ($T^\dagger|+\rangle$) for $d$ even (odd), which is the $(+1)$ eigenstate of $H_{XY}$ ($iZH_{XY}$). Note that initializing the physical ancilla qubit in $T^{(\dagger)}|+\rangle$ is no more difficult than initializing them in any other state.

Compared to Ref.~\cite{chenEfficientMagicState2025}, our deformation circuit is considerably simplified. Ref.~\cite{chenEfficientMagicState2025} took pairs of fold-dual qubits (or pairs of pairs) and concatenated them into a $[[4,2,2]]$ (or $[[6,4,2]]$) error-detecting code. Essentially, they ensured that the fold-dual $X$ and $Z$ operators were sent to overlapping $XX$ and $ZZ$ operators, but also introduced additional self-dual $XXXX$ and $ZZZZ$ stabilizers rather than introducing additional invariant logical degrees of freedom. They required these extra stabilizers in order to maintain the code distance, as pairing fold-dual qubits in the $\mathbb{RP}^2$ code without these additional stabilizers halves the code distance. However, pairing fold-dual qubits in the unrotated surface code does not affect the distance~\cite{chenTransversalLogicalClifford2024}, thus we do not need to introduce additional stabilizers and may use only one ancilla qubit per fold-dual pair and a simpler deformation circuit.

\subsubsection{Measuring logical $H_{XY}$ with a cat ancilla}

\begin{figure*}
    \includegraphics[width=\textwidth]{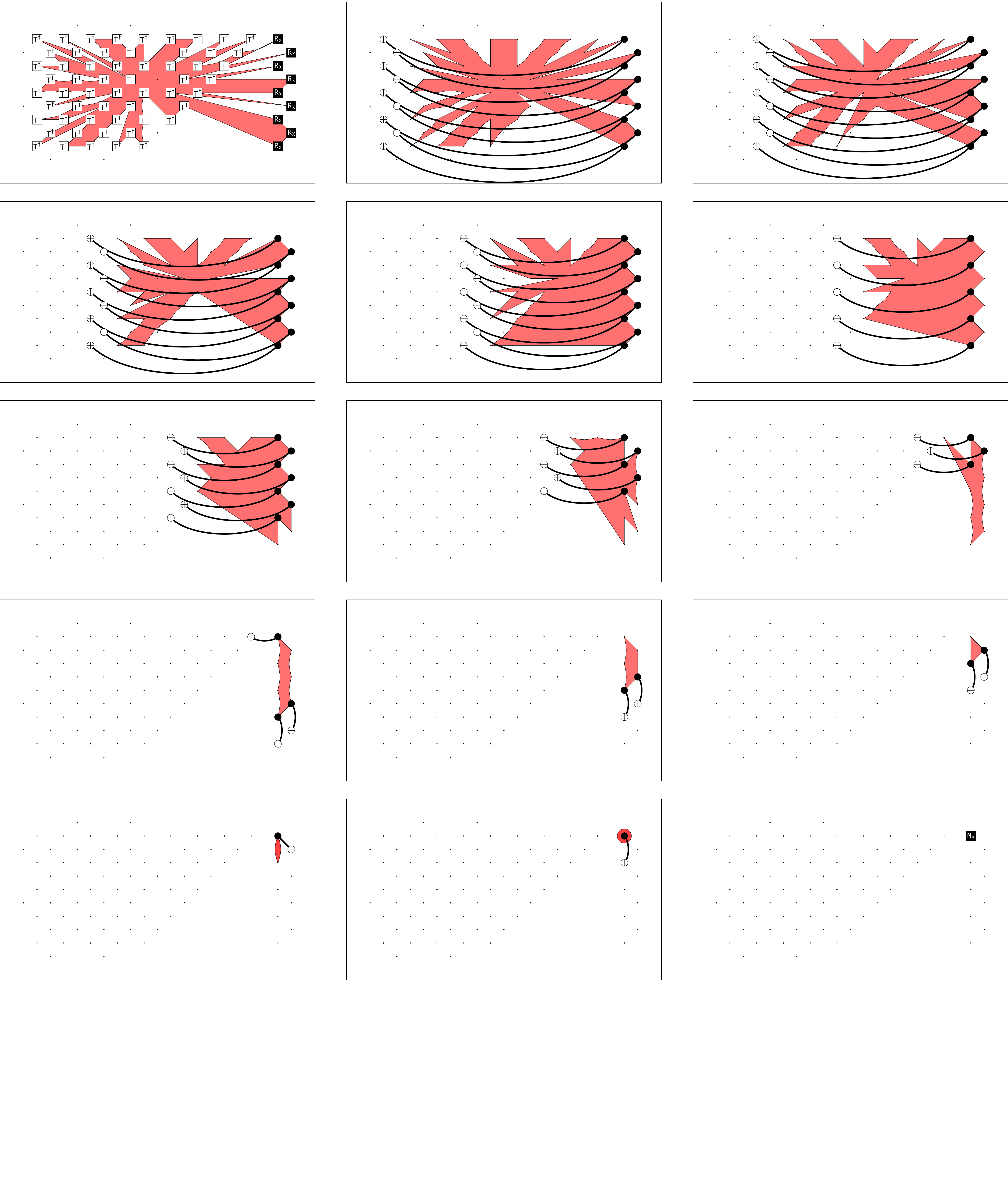}
    \caption{The first half of our double-checking circuit for measuring transversal $H_{XY}$. We begin by applying a layer of $T^\dagger$ gates to the code qubits, to transform the problem into measuring transversal $X$, and initializing a column of ancillas in $|+\rangle$ which will perform the measurement. The large $\otimes X$ operator we are measuring is tracked throughout the circuit. To shrink this operator to a single qubit, we move the column of ancilla qubits across the data qubits, applying $CX$ gates between the ancilla and code qubits at each step, followed by $CX$ gates among the ancilla qubits. Once the operator is supported on a single qubit, we measure that qubit. To complete the double-checking, we run the circuit in reverse and measure each qubit in the ancilla column in the $X$ basis. In the absence of errors, all measurements should be $+1$.}
    \label{fig:CatMeasurement}
\end{figure*}

Once the code has been deformed into a self-dual code, we can measure the logical $H_{XY}$ operator using a cat state ancilla and controlled-$H_{XY}$ gates, as in Refs.~\cite{chenEfficientMagicState2025,chamberlandVeryLowOverhead2020,gidneyMagicStateCultivation2024,itogawaEvenMoreEfficient2024}. Our particular circuit is a ``double-checking'' circuit, as in Ref.~\cite{gidneyMagicStateCultivation2024,chenEfficientMagicState2025} in which the circuit is run in one direction to measure the logical operator, and then run in reverse to measure a set of flag qubits that detect errors during the foreward measurement. We illustrate the foreward half of our double-checking circuit for $d=5$ in Fig.~\ref{fig:CatMeasurement}. For odd $d$ we want to apply a controlled-$H_{XY}$ gate between the ancilla and the code. We convert this to a controlled-$X$ gate by applying a layer of $T^\dagger$ gates, because $T^\dagger H_{XY} T = X$. For even $d$, would instead apply a controlled-$iZH_{XY}$ gate, which we convert to controlled-$X$ gates by applying a layer of $T$ gates, since $T (iZH_{XY}) T^\dagger = X$.

\subsubsection{Reversing the deformation}

After measuring $H_{XY}$, we reverse the deformation circuit and measure the triangular patch of ancillas in the $X$ basis to return to the mid-cycle state, and complete the syndrome extraction cycle to return to the rotated surface code state. If the measurement of the triangular patch of ancilla qubits don't all return $(+1)$, or the final syndrome measurements don't return the correct values, we discard the attempt.

\subsection{Escape}

Our escape stage is simple: We initialize new qubits in $|+\rangle$ and $|0\rangle$ according to standard surface code growth protocols~\cite{liMagicStatesFidelity2015,litinskiGameSurfaceCodes2019,horsman2012surface,fowlerSurfaceCodesPractical2012}, and measure the stabilizers of the larger code several times. We then compute the complementary gap~\cite{gidneyYokedSurfaceCodes2023,gidneyMagicStateCultivation2024} just as in color code cultivation in order to estimate how confident the decoder is in our result. We can postselect on this complementary gap to trade off between the fidelity of the final magic state and the number of retries required to accept the state. 

Note that the most straightforward implementation for computing the complementary gap requires a boundary where logical operators can terminate~\cite{gidneyYokedSurfaceCodes2023,gidneyAnswerHowForce2024}; for this reason, it is not straightforward to compute the complementary gap in $\mathbb{RP}^2$ cultivation. This is why Ref.~\cite{chenEfficientMagicState2025} instead used a different measure (the ``2D soft output"~\cite{meisterEfficientSoftoutputDecoders2024}) to estimate the decoder's confidence, a complication we do not encounter when cultivating directly on the planar surface code.

\section{Conclusion}

We have presented a new surface code cultivation protocol for $|T\rangle$ states that avoids the drawbacks of previous attempts at surface code cultivation. Our simulations and resource estimates show our protocol is competitive with previous color code and $\mathbb{RP}^2$ cultivation protocols when considering uniform depolarizing noise and measures of spacetime volume used in previous work, and likely further improves relative to previous protocols when considering a modified error model without idle errors.

In the future, it will be interesting to consider how these estimates are modified by more realistic models for neutral atom systems. On the optimistic side, the long coherence times of neutral atom qubits means that idle errors are often negligible; we can also incorporate erasure conversion~\cite{wuErasureConversionFaulttolerant2022,sahayHighthresholdCodesNeutralatom2023,maHighfidelityGatesMidcircuit2023,zhangLeveragingErasureErrors2025}, which will allow us to postselect errors more effectively~\cite{jacobyMagicStateInjection2025} and will likely enable the computation of a finer-tuned complementary gap. On the other side, errors induced by qubit movement and trap transfers are not accounted for in our simulations. We also neglect the possibility of leakage, which may be significant for deep cultivation circuits. Leakage is relatively benign in postselected circuits, as it only increases the discard rate; however, if leakage reduction units are required to keep the discard rate reasonable, they may add additional gates and correspondingly increase the noise~\cite{sucharaLeakageSuppressionToric2015,fowlerCopingQubitLeakage2013,mcewenRelaxingHardwareRequirements2023,chowCircuitBasedLeakagetoErasureConversion2024,baranesLeveragingAtomLoss2025}.

We also re-emphasize that beyond the noise model, the measure of spacetime volume in previous works is ad-hoc and not necessarily relevant to neutral atom qubit platforms, More accurate spacetime comparisons will require careful consideration of atom move times and measurement times.

\section*{Note added}
While completing this manuscript, we became aware of two pieces of related work by the Puri group~\cite{sahay2025fold} and researchers at AWS~\cite[updated version of][]{vakninMagicStateCultivation2025}, which appeared in the same arXiv posting as the present work.

\acknowledgements
We thank Jeff Thompson, Sebastian Horvath, and Elmer Guardado-Sanchez for helpful conversations. We thank Kaavya Sahay and Shruti Puri for discussing their related work with us, and coordinating the release of their paper. We thank Ken Brown and Jeff Thompson for suggestions and feedback on the manuscript. The code used to generate noisy STIM circuits and enumerate short-distance errors was modified from code included with~\cite{gidneyMagicStateCultivation2024}.

\appendix

\section*{Appendix: Noise model}
Here, we specify the noise models we use in the simulations. To enable comparison with previous works, we use the same uniform depolarizing noise model of~\cite{gidneyMagicStateCultivation2024,chenEfficientMagicState2025}. The noise model consists of
\begin{itemize}
    \item Single-qubit gates are followed by a single-qubit depolarizing channel with strength $p$.
    \item Two-qubit gates are followed by a two-qubit depolarizing channel with strength $p$.
    \item Idle qubits during single or two-qubit gates incur a single-qubit depolarizing channel with strength $p$.
    \item Initialization in $|0\rangle$ instead prepares $|1\rangle$ with probability $p$, and similar for $|+\rangle$.
    \item Measurement results are flipped with probability $p$.
\end{itemize}
To estimate the advantage of qubits like neutral atoms which may have very long idle coherence times, we also consider a depolarizing noise model in which the idle errors are removed.

\bibliography{CultivationWriteup.bib}

\end{document}